\documentclass[runningheads]{llncs}
\usepackage{graphicx}
\usepackage{hyperref}
\usepackage{booktabs}
\usepackage{sidecap}

\begin{document}
\title{Changes in Policy Preferences in German Tweets during the COVID Pandemic}
\titlerunning{Policy Preferences on Twitter During COVID}
% If the paper title is too long for the running head, you can set
% an abbreviated paper title here
%

\author{Felix Biessmann\inst{1,2}
% \orcidID{0000-0002-3422-1026}
} 
% \and
% Second Author\inst{2,3}\orcidID{1111-2222-3333-4444}
%
\authorrunning{F. Biessmann}
% First names are abbreviated in the running head.
% If there are more than two authors, 'et al.' is used.
%
\institute{Berlin University of Applied Sciences \and
Einstein Center Digital Future, Berlin, Germany \\
\email{felix.biessmann@bht-berlin.de}
}

\maketitle              % typeset the header of the contribution
\begin{abstract}
Online social media have become an important forum for exchanging political opinions. In response to COVID measures citizens expressed their policy preferences directly on these platforms. Quantifying political preferences in online social media remains challenging: The vast amount of content requires scalable automated extraction of political preferences -- however fine grained political preference extraction is difficult with current machine learning (ML) technology, due to the lack of data sets. Here we present a novel data set of tweets with fine grained political preference annotations. A text classification model trained on this data is used to extract policy preferences in a German Twitter corpus ranging from 2019 to 2022. Our results indicate that in response to the COVID pandemic, expression of political opinions increased. Using a well established taxonomy of policy preferences we analyse fine grained political views and highlight changes in distinct political categories. These analyses suggest that the increase in policy preference expression is dominated by the categories pro-welfare, pro-education and pro-governmental administration efficiency. All training data and code used in this study are made publicly available to encourage other researchers to further improve automated policy preference extraction methods. We hope that our findings contribute to a better understanding of political statements in online social media and to a better assessment of how COVID measures impact political preferences. 
% The policy preferences extracted with such models could help to balance the political spectrum of social network texts presented to users via automated feeds and ultimately encourage users to consume less biased political information. 

\end{abstract}
\section{Introduction}
The past decades have shown two trends that are becoming increasingly interdependent: Political campaigns take place online in social media. And at the same time online content for individual users is recommended using automated machine learning (ML) systems that are often optimized for user engagement or other proxy metrics for economic profit. These mechanisms can increase visibility of polarizing content and simultaneously enforce a bias towards existing user preferences. 

% Some of the resulting negative consequences include the increased visibility of hate speech and political radicalization. These problems have been recognized also by the research community and there are substantial efforts to develop data sets and models to detect violent content in online social media \cite{macavaney2019hate}. \\

During the COVID pandemic, global platforms such as online social media allowed users to directly express their preferences for or against the measures taken by governments, such as lockdowns or vaccination programs. Analysing these policy preferences can yield valuable insights that could help to improve governmental policies. 
The large amount of content requires methods for automated extraction of policy preferences. Recent trends in machine learning (ML) towards bigger and more powerful language models could help to improve policy preference extraction. However there are few training data sets that contain annotations for fine grained policy preferences \cite{doan2022survey}. 
The lack of high quality annotated data sets with political information impedes the development of better models for automated detection of policy preferences. \\

%
% Without such models however it is difficult to analyse improve the balance of ML based content recommenders. In order to reduce bias for a certain political direction in existing recommenders, we need a better understanding of how policy preferences are expressed in online content. 
%
Here we present a data set of online social media content, Twitter posts, with fine grained political annotations as defined in \cite{manifesto_coding_scheme}. The data set is used to train a text classification model that predicts policy preferences from social network posts. On a larger corpus of tweets collected from 2019 to 2022 the model is used to predict policy preferences before and during the COVID pandemic. 
Analyses of automatically extracted policy preferences suggest that the amount of policy preferences expressed on Twitter increased after the first lockdown. Leveraging a fine grained political viewpoint taxonoomy we can investigate which policy preferences were expressed in those political tweets. 
To summarize, the main contributions of this study are:

\begin{itemize}
    \item A data set of German tweets with fine grained political preference annotation
    \item A novel text classification model
    \item An analysis of policy preferences before and during the COVID pandemic
\end{itemize}

% In collaboration with the Manifesto Project we trained annotators to label thousands of tweets according to the Manifesto Project coding scheme \cite{manifesto_coding_scheme}. 

\section{Related Work}

The general topic of automated information extraction from online socia media has been widely studied and different approaches have been proposed, including supervised ML methods, such as text classification \cite{gryc_leveraging_2014}, and unsupervised methods, such as topic models, or extensions thereof \cite{aiello_sensing_2013,biessmann_quantifying_2012,biesmann_canonical_2012}. Many of these methods are dedicated to trending topic extraction. Since not all trending topics are related to the political discourse a large fraction of these methods do not lend themselves easily to the invesigation of policy preferences. 

A number of studies have explored automated extraction of policy preferences, for a comprehensive overview we refer the interested reader to \cite{doan2022survey}. There have been many studies exploring traditional ML techniques for ideology detection and policy preference extraction \cite{thomas_get_2012} as well as approaches based on more recent natural language processing models, such as Recurrent Neural Networks \cite{iyyer_political_2014} or more recently also Transformers \cite{schick_exploiting_2021}. 

The authors of \cite{doan2022survey} highlight that training ML models for automated extraction of fine grained policy preferences expressed in online social media content remains challenging. Primarily this is due to the fact that annotating this data requires expertise that can not as easily be crowdsourced, as the annotation of hate speech for instance. Annotation of policy preferences requires domain expertise and in particular experience with policy preferences as expressed in online media. 

There are some publicly available data sets that can be used for training ML models that detect policy preferences in text data. One of the largest and best curated data sets is the corpus of the Manifesto Project \cite{Volkens:2021} which contains over 1,500,000 quasi-sentences, extracted from over 1,500 party manifestos, and annotated according to a well established category scheme of 56 policy categories \cite{manifesto_coding_scheme}. 
This data has been used by researchers to investigate policy preferences \cite{krause2020appearing} and there have been efforts to train ML models on this data to make predictions on online social media texts \cite{biessmann2016predicting,schwarz2019detecting,phillips2020party}. However the texts of party manifestos are written in a different style than posts in online social media. Hence models trained on the manifesto data usually do not work well on online social media texts. 
Other data sets focus more on texts in online social media but these often focus on a small set of political policy preferences \cite{DVN/F9ICHH_2015,kiesel2019semeval,aksenov2021fine,DVN/BKGZUL_2021}.

% \begin{minipage}{.45\textwidth}
\begin{table}
\footnotesize
\centering
\begin{tabular}{lrrrl}
\toprule
{} &  precision &  recall &  f1-score &   support \\
\midrule
controlled economy +   &  1.00 &  0.67 &  0.80 &  3.0 \\
europe -               &  0.80 &  0.75 &  0.77 &  16.0 \\
environmentalism +     &  0.76 &  0.70 &  0.73 &  90.0 \\
democracy +            &  0.63 &  0.74 &  0.68 &  77.0 \\
anti-imperialism +     &  1.00 &  0.50 &  0.67 &  2.0 \\
economic orthodoxy +   &  0.57 &  0.67 &  0.62 &  6.0 \\
europe +               &  0.56 &  0.64 &  0.60 &  14.0 \\
undefined              &  0.58 &  0.55 &  0.57 &  271.0 \\
infrastructure +       &  0.43 &  0.80 &  0.56 &  20.0 \\
foreign special +      &  0.50 &  0.55 &  0.52 &  11.0 \\
\dots &&&&\\
\midrule
accuracy               &   &   &   0.46 & 1214  \\
macro avg              &  0.30 &  0.31 &  0.30 &  1214 \\
weighted avg           &  0.46 &  0.46 &  0.46 &  1214 \\
\bottomrule
\end{tabular}
\label{tab:results_short}
\caption{F1 scores for tweets in the test set for the top 10 (according to F1) political categories. The complete list can be found in the Appendix, table \ref{tab:results}.}
\end{table}
% \end{minipage}

% \section{PoliTweets Training Data Set}
\section{Training Data Set}
\label{sec:training_data}
For annotating training data with fine grained policy preferences we sampled tweets from a corpus of German tweets \cite{nane_kratzke_2020_3633935}. 
The tweets 
% in our \textit{PoliTweets} data set 
were sampled between August 2019 and March 2022 and filtered using the following criteria:

\paragraph{User Interaction} We selected tweets that were interacted with in some form (likes, retweets, quotes) at least once.

\paragraph{Relevance} We used a ML model (see below) trained on the Manifesto Project corpus \cite{Volkens:2021} to estimate the political relevance of each tweet. To increase the usefulness of the annotated data set we tried to cover all labels of the Manifesto Project's category scheme by selecting for each week only the top 5 tweets that were predicted as the most likely for each political category by an ML model trained on German party manifestos \cite{Volkens:2021}. 

The filtered set of tweets were then annotated by two experts trained by researchers of the Manifesto Project. The annotation was performed in a custom written web app and later using labelstudio \cite{lstudio}. Annotators were instructed to label a tweet with one of the 56 political categories of the Manifesto Project codebook \cite{manifesto_coding_scheme}. 

Additionally annotators provided the label \texttt{undefined} for tweets that could not be associated with any of the relevant political categories. If the tweet contained an image, annotators also considered the image content for the annotation. Context beyond the actual tweet content was not taken into account. Exceptions were tweets that replied to or commented on another tweet. In that case the original tweet was also considered. These replied-to tweets are, to keep the data set simpler, not part of the data set but can be retrieved via the url of the annotated tweet.  

In the current version of the data set there are 6097 unique tweets and the most frequent political categories annotated are shown in \autoref{tab:class_histogram} (Appendix). Note that the majority of tweets is labeled as \texttt{undefined}, despite the filtering with the ML model. This is an indication that the data set contains useful negative examples for training better models. The data set is released and available for research purposes \cite{politweets}.

\begin{figure}
    \centering
    \includegraphics[width=4cm]{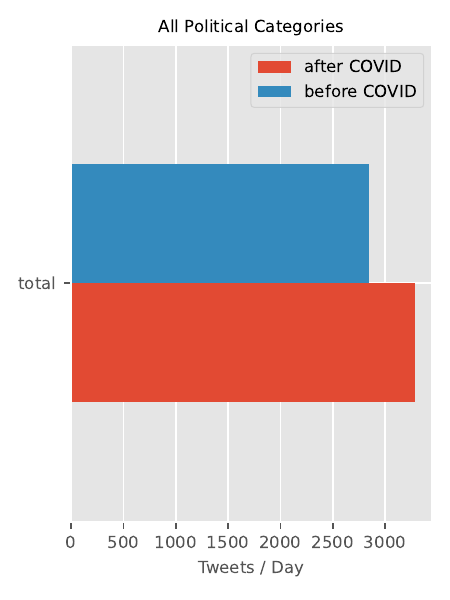}
    \includegraphics[width=4cm]{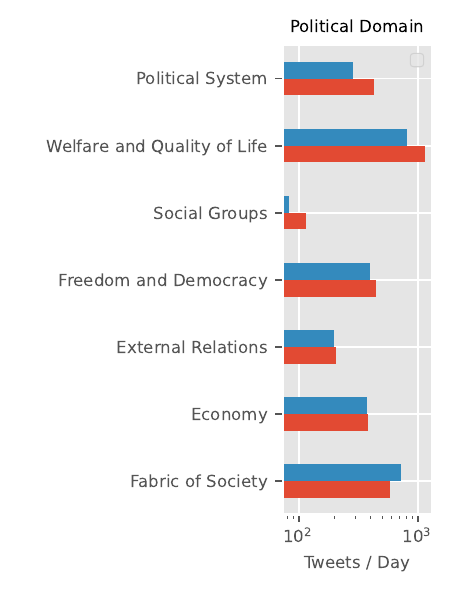}
    \includegraphics[width=4cm]{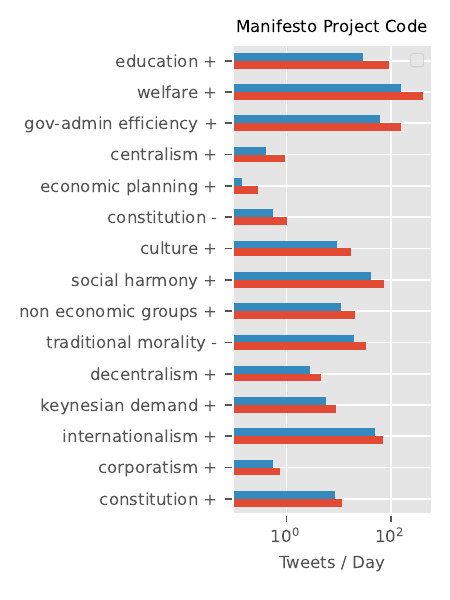}
    \caption{Increases in political tweets after the first COVID lockdown in Germany. Policy preferences were extracted with a text classifier. \textit{Left:} After the first lockdown the total number of political tweets per day increases. \textit{Middle:} Strong increases were observed in the broad political category of \texttt{political system} and \texttt{welfare}; note the log scale on the x-axis. \textit{Right:} Fine grained policy preferences show a strong increase in \textit{pro education, pro welfare} and \textit{pro government administration efficiency}}
    \label{fig:increases}
\end{figure}

\section{Evaluation of Policy Preference Predictors}
\label{sec:model}
To establish a simple baseline for policy preference extraction on the PoliTweet data set we used the TextPredictor module of the autoML package AutoGluon \cite{autogluon,agmultimodaltext}. The model was trained on a V100 NVIDIA GPU with a pretrained BERT model checkpoint (\texttt{bert-base-german-cased}) on the entire German part of the manifesto corpus \cite{Volkens:2021} and 4883 annotated tweets from the training data set in
% PoliTweets data
\autoref{sec:training_data}; 1214 annotated tweets were held out for testing the model. In \autoref{tab:results_short} we list the results for the top 10 political categories that could be predicted with highest F1 score by the model; the full list of results for all categories is listed in the Appendix, \autoref{tab:results}. Note that while the overall prediction performance is below 0.50 F1 score (macro or class-frequency weighted), these results are still encouraging. Fine grained political viewpoint extraction is a challenging task and even when trained on the manifesto corpus, the classification performance for all categories with extensive tuning and leveraging state-of-the-art ML models often stays below an F1 score of 0.5 \cite{subramanian_hierarchical_2018}.
\section{Policy Preferences after COVID lockdown}
The model as trained in \autoref{sec:model} was then applied to the entire twitter corpus \cite{nane_kratzke_2020_3633935} between 2019 and 2022 and filtered using the relevance and activity criteria as mentioned in \autoref{sec:training_data}. We applied additional relevance filters to extract only tweets expressing political views. All tweets for which the category \texttt{undefined} was amongst the top 3 predictions of the text classification model were considered irrelevant and filtered out. The histograms of policy preferences in the remaining tweets were then compared before and after the COVID lockdown onset in Germany. In \autoref{fig:increases} we show histograms of political views expressed in tweets before and after onset of the first lockdown. 

Overall our results suggest that the number of political tweets increased after the first lockdown. 
\begin{SCfigure}
    \centering
    \caption{Number of tweets over time in those political categories that exhibit a strong increase after the first lockdown in Germany. Bottom panel shows an overview of COVID cases reported from Robert-Koch-Institutel, lockdown starts are indicated in blue. While the first lockdown did not result in strong increases of Tweets with political preferences, during the second COVID wave political preferences in the categories \textit{pro education, pro welfare} and \textit{pro government administration efficiency} were expressed more often than before.  \label{fig:timeline}\vspace{2cm}}
    \includegraphics[width=.55\textwidth]{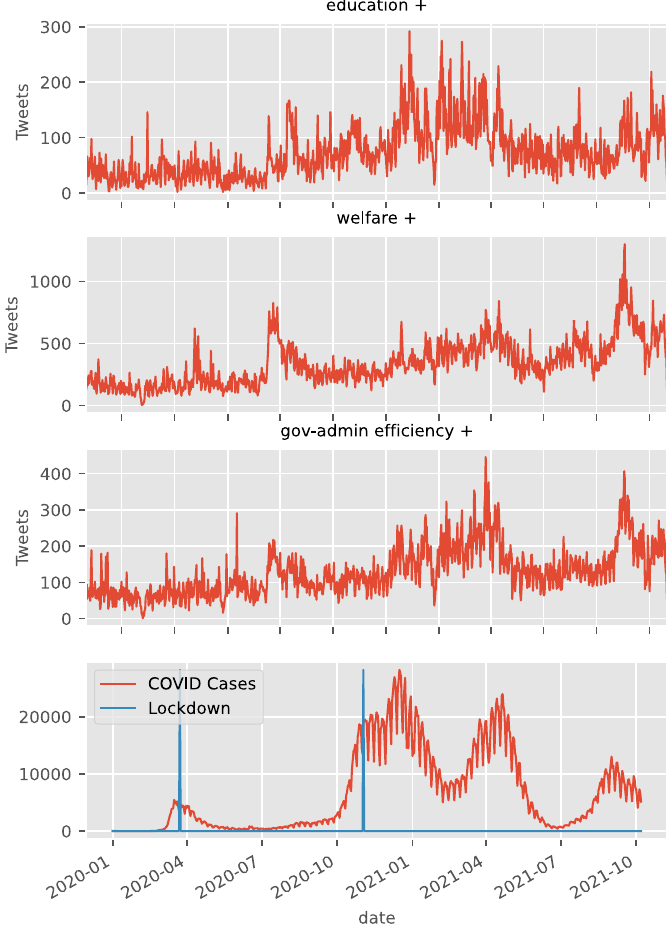}
\end{SCfigure}
Investigating the fine grained political categories we find that this increase is driven by an increased number of tweets categorized as \textit{pro education, pro welfare} and \textit{pro government administration efficiency}. These changes in policy preferences of tweets could reflect the negative impact that COVID measures such as lockdowns had: many employes lost their jobs, many needed to teach their children at home and all administrational processes were substantially slowed down due to the poor digitalization in German administration. 

In \autoref{fig:timeline} timelines are shown for the political categories \textit{pro education, pro welfare} and \textit{pro government administration efficiency}, which exhibit the largest change after onset of the COVID lockdown as shown in \autoref{fig:increases}. The bottom panel in \autoref{fig:timeline} shows the onsets of lockdowns and COVID case numbers. The strongest impact of lockdown measures with respect to political policy preferences on Twitter appears to develop during the second wave of the pandemic.

\section{Conclusion}
This study presents three main contributions, a) a data set of German tweets with fine grained political preference annotation, b) a novel text classification model trained on that data and c) an analysis of policy preferences before and during the COVID pandemic. Our preliminary analyses of tweets during the COVID pandemic showed a pronounced increase in political tweets overall and in particular also in certain fine grained political categories. These findings are not from a representative sample and have several other limitations, such as the predictive performance of the text classification model for some classes, especially in rare categories. Nonetheless we believe the data set, the model and the experimental results are an important step towards more scalable and more fine grained policy preference extraction in German online social media. We hope that our contributions will encourage other researchers to improve current state-of-the-art models for policy preference extraction in research and applications.
% , also for other application domains, such as news recommender systems, and for other researchers from social sciences.

% \clearpage
\section*{Acknowledgements}
We thank Jonas Bauer for conceptualizing, implementing and maintaining the first data annotation setup, Teo Chiaburu for setting up labelstudio, Marvin Müller and Maren Krumbein for annotating tweets, Pola Lehmann for training the annotators and valuable feedback on the analyses, Johannes Hoster for analyses and Philipp Staab for valuable discussions on sociological aspects.

\scriptsize
\bibliographystyle{splncs04}
\bibliography{politweets.bib}
%
% ---- Bibliography ----
%
% BibTeX users should specify bibliography style 'splncs04'.
% References will then be sorted and formatted in the correct style.
%
% \bibliographystyle{splncs04}
% \bibliography{mybibliography}
%

\begin{table}
\scriptsize
\centering
\begin{tabular}{lr}
\toprule
Political Category &  Count \\
\midrule
undefined              &                  1318 \\
freedom/human rights + &                   502 \\
environmentalism +     &                   401 \\
democracy +            &                   395 \\
social justice +       &                   379 \\
welfare +              &                   343 \\
political authority +  &                   337 \\
national way of life + &                   241 \\
national way of life - &                   198 \\
infrastructure +       &                   136 \\
social harmony +       &                   133 \\
gov-admin efficiency + &                   131 \\
labour +               &                   130 \\
education +            &                   119 \\
law and order +        &                   112 \\
free enterprise +      &                   104 \\
multiculturalism -     &                    94 \\
europe +               &                    90 \\
europe -               &                    63 \\
political corruption - &                    60 \\
anti-growth economy +  &                    57 \\
internationalism +     &                    54 \\
multiculturalism +     &                    49 \\
constitution +         &                    47 \\
traditional morality + &                    43 \\
military -             &                    43 \\
traditional morality - &                    42 \\
market regulation +    &                    41 \\
productivity +         &                    41 \\
military +             &                    37 \\
foreign special +      &                    33 \\
agriculture +          &                    33 \\
welfare -              &                    31 \\
economic orthodoxy +   &                    31 \\
culture +              &                    28 \\
marxist analysis +     &                    25 \\
economic goals         &                    25 \\
non economic groups +  &                    19 \\
peace +                &                    15 \\
incentives +           &                    15 \\
controlled economy +   &                    13 \\
nationalization +      &                    12 \\
protectionism +        &                    10 \\
keynesian demand +     &                    10 \\
internationalism -     &                    10 \\
anti-imperialism +     &                    10 \\
decentralism +         &                     9 \\
centralism +           &                     5 \\
middle class +         &                     5 \\
foreign special -      &                     5 \\
protectionism -        &                     4 \\
labour -               &                     3 \\
minority groups +      &                     2 \\
education -            &                     2 \\
economic planning +    &                     1 \\
constitution -         &                     1 \\
\midrule
total & 6097\\
\bottomrule
\end{tabular}
\caption{Histogram of all labels annotated in the PoliTweet data set according to the Manifesto Project political category taxonomy. }
\label{tab:class_histogram}
\end{table}

\begin{table}
\centering
\scriptsize
\begin{tabular}{lrrrl}
\toprule
{} &  precision &  recall &  f1-score &   support \\
\midrule
controlled economy +   &  1.00 &  0.67 &  0.80 &  3.0 \\
europe -               &  0.80 &  0.75 &  0.77 &  16.0 \\
environmentalism +     &  0.76 &  0.70 &  0.73 &  90.0 \\
democracy +            &  0.63 &  0.74 &  0.68 &  77.0 \\
anti-imperialism +     &  1.00 &  0.50 &  0.67 &  2.0 \\
economic orthodoxy +   &  0.57 &  0.67 &  0.62 &  6.0 \\
europe +               &  0.56 &  0.64 &  0.60 &  14.0 \\
undefined              &  0.58 &  0.55 &  0.57 &  271.0 \\
infrastructure +       &  0.43 &  0.80 &  0.56 &  20.0 \\
foreign special +      &  0.50 &  0.55 &  0.52 &  11.0 \\
education +            &  0.38 &  0.76 &  0.51 &  17.0 \\
marxist analysis +     &  0.50 &  0.50 &  0.50 &  6.0 \\
social justice +       &  0.46 &  0.52 &  0.49 &  69.0 \\
internationalism +     &  0.39 &  0.50 &  0.44 &  14.0 \\
freedom/human rights + &  0.45 &  0.43 &  0.44 &  97.0 \\
political authority +  &  0.41 &  0.43 &  0.42 &  60.0 \\
decentralism +         &  0.33 &  0.50 &  0.40 &  2.0 \\
traditional morality - &  0.43 &  0.38 &  0.40 &  8.0 \\
protectionism +        &  0.33 &  0.50 &  0.40 &  2.0 \\
law and order +        &  0.37 &  0.42 &  0.39 &  24.0 \\
social harmony +       &  0.42 &  0.35 &  0.38 &  23.0 \\
political corruption - &  0.36 &  0.38 &  0.37 &  13.0 \\
agriculture +          &  0.50 &  0.29 &  0.36 &  7.0 \\
constitution +         &  0.33 &  0.40 &  0.36 &  10.0 \\
national way of life - &  0.33 &  0.36 &  0.35 &  36.0 \\
multiculturalism -     &  0.38 &  0.33 &  0.35 &  18.0 \\
culture +              &  0.29 &  0.40 &  0.33 &  5.0 \\
military -             &  0.33 &  0.31 &  0.32 &  13.0 \\
macro avg              &  0.30 &  0.31 &  0.30 &  1214.0 \\
welfare +              &  0.37 &  0.24 &  0.29 &  82.0 \\
market regulation +    &  0.33 &  0.25 &  0.29 &  12.0 \\
national way of life + &  0.30 &  0.29 &  0.29 &  48.0 \\
military +             &  0.22 &  0.33 &  0.27 &  6.0 \\
peace +                &  0.17 &  0.33 &  0.22 &  3.0 \\
labour +               &  0.25 &  0.19 &  0.22 &  21.0 \\
gov-admin efficiency + &  0.21 &  0.19 &  0.20 &  32.0 \\
free enterprise +      &  0.19 &  0.15 &  0.17 &  20.0 \\
protectionism -        &  0.00 &  0.00 &  0.00 &  2.0 \\
centralism +           &  0.00 &  0.00 &  0.00 &  1.0 \\
welfare -              &  0.00 &  0.00 &  0.00 &  8.0 \\
traditional morality + &  0.00 &  0.00 &  0.00 &  9.0 \\
corporatism +          &  0.00 &  0.00 &  0.00 &  0.0 \\
incentives +           &  0.00 &  0.00 &  0.00 &  2.0 \\
economic goals         &  0.00 &  0.00 &  0.00 &  8.0 \\
productivity +         &  0.00 &  0.00 &  0.00 &  6.0 \\
education -            &  0.00 &  0.00 &  0.00 &  1.0 \\
nationalization +      &  0.00 &  0.00 &  0.00 &  0.0 \\
multiculturalism +     &  0.00 &  0.00 &  0.00 &  4.0 \\
minority groups +      &  0.00 &  0.00 &  0.00 &  0.0 \\
foreign special -      &  0.00 &  0.00 &  0.00 &  1.0 \\
anti-growth economy +  &  0.00 &  0.00 &  0.00 &  9.0 \\
keynesian demand +     &  0.00 &  0.00 &  0.00 &  1.0 \\
internationalism -     &  0.00 &  0.00 &  0.00 &  3.0 \\
non economic groups +  &  0.00 &  0.00 &  0.00 &  1.0 \\
\midrule
accuracy               &   &   &   0.46 & 1214  \\
macro avg              &  0.30 &  0.31 &  0.30 &  1214 \\
weighted avg           &  0.46 &  0.46 &  0.46 &  1214 \\
\bottomrule
\end{tabular}
\caption{F1 scores for tweets of all political categories in the test set.}
\label{tab:results}
\end{table}

\end{document}